\documentclass[a4paper,fleqn]{cas-sc}

\usepackage[authoryear,longnamesfirst,numbers]{natbib}
\usepackage{amsmath,amsfonts}
\usepackage{url}
\usepackage{algpseudocode}
\usepackage{algorithm}
\usepackage{xcolor}
\usepackage{booktabs}
\usepackage{braket}
\usepackage{cleveref}
\crefname{equation}{}{}
\Crefname{equation}{Equation}{Equations}
\crefname{figure}{Fig.}{Figs.}
\crefname{table}{Table}{Tables}
\crefname{section}{Section}{Sections}
\crefname{algorithm}{Algorithm}{Algorithms}
\usepackage{float}
\usepackage{lineno}

\begin{document}

\shorttitle{A Framework for Solving Continuous Problems using AQC}    
\shortauthors{Z. Kaseb et al.}  

\title [mode = title]{A Framework for Solving Continuous Energy and Power System Problems using Adiabatic Quantum Computing}  

\author[1]{Zeynab Kaseb}[orcid=0000-0002-5142-290]

\cormark[1]
\ead{Z.Kaseb@tudelft.nl}
\author[2]{Matthias M\"oller}
\author[1]{Peter Palensky}
\author[1]{Pedro P. Vergara}

\affiliation[1]{organization={Electrical Sustainable Energy, Delft University of Technology, The Netherlands}}
\affiliation[2]{organization={Applied Mathematics, Delft University of Technology, The Netherlands}}
\cortext[1]{Corresponding author}

\begin{abstract}
The increasing scale and nonlinearity of modern energy and power system problems pose significant challenges to classical numerical solvers. In parallel, advances in quantum and quantum-inspired hardware are expected to improve scalability and offer performance advantages for large-scale optimization problems. Therefore, we propose a novel combinatorial optimization framework that reformulates continuous energy and power system problems into a format executable on quantum/digital annealers. The proposed framework accommodates both real and complex numbers and can represent both linear and nonlinear equations. As a proof of concept, we demonstrate its use in three applications: (i) 2D steady conductive heat transfer for a plate with constant temperature at each edge, where coefficient and boundary condition matrices are developed to solve linear system of equations, (ii) power system parameter identification, where the admittance matrix is estimated given voltage and current measurements, and (iii) power flow analysis, which solves the governing equations for active and reactive power balance. As a proof of concept, the applications are run on small test cases. The results show that the framework effectively and efficiently addresses the three applications and therefore suggest its potential to solve a wide range of energy and power system problems.
\end{abstract}



\maketitle


\section{Introduction}\label{sec: introduction}
\noindent Energy and power systems are governed by a wide range of mathematical models describing thermal, electrical, mechanical, and electromagnetic processes across multiple temporal and spatial scales \cite{Brenan1995,Jin2024}. Such models arise in applications including heat conduction, computational fluid dynamics, electromagnetic transient simulation, power flow and optimal power flow problems, dynamic stability assessment, state estimation, and parameter identification, among others \cite{Milano2010,JiyuanTu2013}. Despite their diversity, these applications share a common computational characteristic, that is, they require solving systems of coupled equations involving continuous variables, which may be algebraic or differential, linear or nonlinear, and defined over either real or complex domains \cite{Quarteroni2007}. 

Modeling such systems of equations is increasingly challenging as modern energy infrastructures become more interconnected, decentralized, and heterogeneous. Renewable integration, distributed energy resources, sector coupling, and multi-physics interactions have significantly increased model dimensionality and nonlinearity \cite{Saad2003}. In many cases, the primary difficulty is no longer only deriving physically accurate governing equations, but also constructing efficient computational representations of those equations that remain scalable and numerically tractable as complexity grows. Classical numerical approaches, including Newton-based iterative solvers, sparse linear algebra methods, and gradient-based nonlinear optimization, remain the backbone of energy-system computation. Yet their performance can degrade in the presence of ill-conditioned Jacobians, nonconvex nonlinearities, stiff differential-algebraic formulations, poor initialization, weak observability, or large-scale coupled models \cite{JorgeNocedal2006,Soares2018}.

At the same time, quantum annealers and quantum-inspired digital annealers have emerged as alternative computing platforms for combinatorial optimization \cite{Jaschke2019}. These architectures are designed to solve discrete binary optimization problems, typically expressed in Ising model or quadratic unconstrained binary optimization (QUBO) form \cite{Nutricati2025}. Their potential lies in exploring large combinatorial spaces efficiently through hardware-level parallelism \cite{Morstyn2023,Feng2021}. However, a fundamental mismatch exists between these machines and most physical system models, which are naturally continuous, whereas annealers operate on discrete binary variables. This gap makes model reformulation the central challenge in applying annealing technologies to practical problems \cite{Zhang2025}. 

Therefore, this paper proposes a combinatorial optimization framework for modeling continuous energy and power system problems on Ising machines. The framework reformulates mathematical models, including linear and nonlinear equations, real- and complex-valued variables, and problems derived from both algebraic and differential governing equations once discretized, into a discrete optimization problem, which is solvable by quantum or digital annealers\footnote{\url{https://www.fujitsu.com/global/services/business-services/digital-annealer}}. We have initiated the use of this approach for PF analysis and optimal power flow (OPF) and performed simulations for test systems up to 1354 buses~\cite{Kaseb2024power,Kaseb2025enhanced}. Our work has further established tolerance to missing bus information through partitioned formulations~\cite{Kaseb2024power}, addressed ill-conditioned scenarios~\cite{Kaseb2025enhanced}, and proved applicability in a hardware-in-the-loop setting for real-time simulation~\cite{Kaseb2025QHIL}.

In the present study, we extend the framework to highlight its generalization ability beyond PF analysis and OPF. We demonstrate that the same framework can be applied to other problems, provided they can be formulated as root-finding tasks. To do so, three fundamentally different applications are considered in this paper: (i) a 2D steady-state conductive heat transfer problem demonstrates the treatment of linear real-valued systems arising from the discretization of partial differential equations, (ii) a power system parameter identification problem demonstrates inverse estimation of a complex-valued admittance matrix from measurement data, and (iii) a nonlinear power flow analysis demonstrates solving coupled nonlinear complex-valued equations describing active and reactive power balance. As a proof of concept, the first application is demonstrated on a $4 \times 4$ plate, and the results are compared with those of a classical linear solver. The remaining two applications are quantitatively validated on a standard 4-bus test system~\cite{Grainger1994}, and the results are compared against those of the NR solver implemented in Pandapower~\cite{thurner2018}. 

The experiments are conducted using D-Wave's Advantage\texttrademark\, quantum annealer (QA)\footnote{\url{https://www.dwavesys.com}}, and Fujitsu's Quantum-Inspired Integrated Optimization software (QIIO)\footnote{\url{https://en-portal.research.global.fujitsu.com/kozuchi}}, which emulates simulated annealing by massively parallel execution of Markov Chain Monte Carlo processes and offers higher qubit connectivity than QA~\cite{Yin2023}. The small test system size is limited by our access to hardware and the prohibitive computational cost of large-scale simulations. Note that although current quantum hardware has limitations, recent studies have shown that adiabatic quantum computing is already viable for addressing certain industrial-scale problems~\cite{Yarkoni2022}. QA is a representative noisy intermediate-scale quantum (NISQ) hardware system, and our proposed framework is explicitly designed to operate within the constraints and opportunities of such hardware. This work, therefore, not only lays essential groundwork for leveraging future advancements in quantum technology but also provides practical formulations that align with the current capabilities of quantum annealers. Furthermore, the proposed framework can operate both as a standalone solver and as a hybrid component, such as providing a warm start to the classical NR solver to potentially enhance the convergence behavior.

\section{Methodology}
\noindent Mathematical models describing continuous energy and power system problems can be reformulated into discrete and/or combinatorial optimization problems that are directly compatible with quantum and digital annealers. In our framework, the reformulation is expressed as a quadratic unconstrained binary optimization (QUBO), a standard mathematical representation that serves as the computational backbone of adiabatic quantum computing. This QUBO representation is then mapped to a problem Hamiltonian, which defines the search space explored by quantum and digital annealers. An overview of the proposed framework is shown in~\cref{fig:diagram}. The involved components of the framework are described in \cref{sec:rootfinding,sec:discretization,sec:combinatorialformulation,sec:orderreduction,sec:qubosolving,sec:deltaupdate}:

\begin{figure}[h]
    \centering
\includegraphics[width=3in]{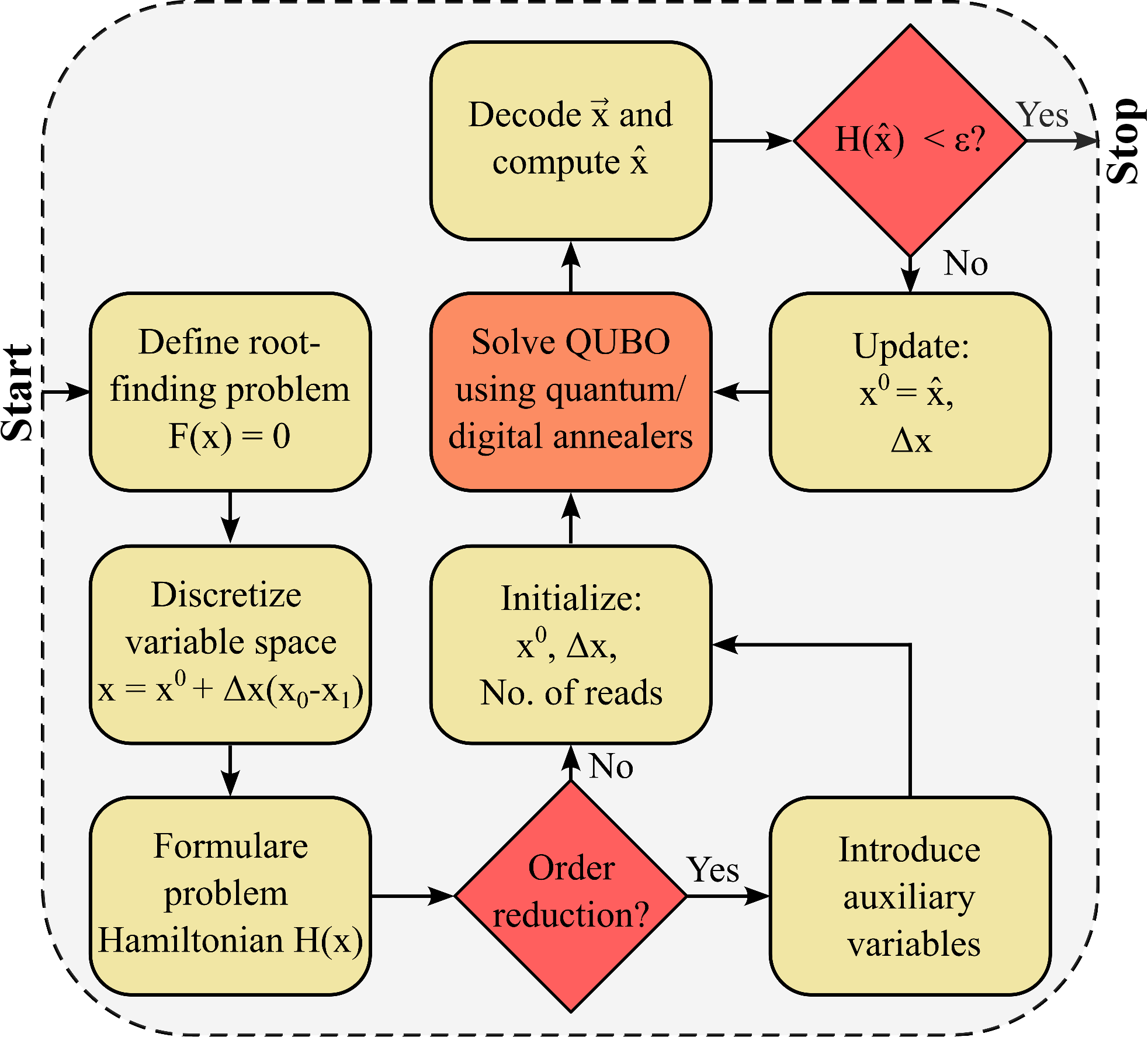}
    \caption{Schematic representation of the proposed framework for solving energy and power system problems with Ising machines, e.g., quantum and digital annealers.}
    \label{fig:diagram}
\end{figure}

\subsection{Define root-finding problem}\label{sec:rootfinding}
\noindent The first step is to define the problem as a root-finding problem of the form:
\begin{equation}
    F(x) = 0, \quad x \in \mathbb{R},
    \label{eq:root-finding=problem}
\end{equation}
where \( F(x) \) may consist of linear and nonlinear terms. For illustration purposes, we describe the methodology for a scalar variable and apply it to vector-valued variables in \cref{sec:HT,sec:LinProblem,sec:NonLinProblem}.

\subsection{Discretize variable space}\label{sec:discretization}
\noindent The continuous variables (\(x\)) must then be discretized \((\hat{x}\)) to transform the problem into a combinatorial optimization problem. A straightforward discretization scheme can be expressed as:
\begin{equation}
    \hat{x} = x^0 + \Delta x (x_{0} - x_{1}),
    \label{eq:update}
\end{equation}
where \( x^0\in\mathbb{R} \) is a base value, \( \Delta x\in\mathbb{R} \) represents the step size, and \( x_{0}, x_{1} \in \{0,1\} \) are \emph{binary decision variables} whose value decides whether the base value $x^0$ is increased ($x_0 = 1 \,\land\, x_1 = 0$), decreased ($x_0 = 0 \,\land\, x_1 = 1$), or kept at its current value ($x_0 = 0 \,\land\, x_1 = 0$ or $x_0 = 1 \,\land\, x_1 = 1$).

\subsection{Formulate combinatorial optimization problem}\label{sec:combinatorialformulation}
\noindent An alternative to solving the root-finding problem \eqref{eq:root-finding=problem} directly is to square the residual function and solve the combinatorial optimization problem for \(\mathbf{x}=\{x_{0}, x_{1}\}\):
\begin{equation} \label{eq:hamiltonian}
    \min_{\mathbf{x}\in\{0,1\}^2} \sum F^2(\hat{x}).
\end{equation}

If \( F(x) \) contains only first-order terms, then \( F^2(\hat{x}) \) expands into a function consisting of constant, linear, and quadratic terms. These correspond to the \emph{offset}, \emph{off-diagonal}, and \emph{diagonal} terms of the QUBO matrix, respectively. A general QUBO problem is defined as:
\begin{equation}
        \min_{\mathbf{x} \in \{0,1\}^n} \sum_{i=1}^{n} \sum_{k=1}^{n} Q_{ik} x_i x_k,
    \label{eq:qubo}
\end{equation}
where \( Q \in \mathbb{R}^{n \times n} \) is a symmetric, real-valued matrix encoding the quadratic and linear coefficients. Note that the \emph{offset} term represents a constant shift in the objective function that changes the value of the optimal solution to the minimization problem \eqref{eq:qubo} but not the optimal bitstring \( \mathbf{x} \in \{0,1\}^n \).

\subsection{Order reduction}\label{sec:orderreduction}
\noindent In cases where the original function \( F(x) \) contains higher-order terms, the squared function \( F(\hat{x})^2 \) introduces terms beyond the quadratic form. To preserve the QUBO structure \eqref{eq:qubo}, a higher-order reduction technique is employed that introduces auxiliary binary variables \( z \) such that \(x_i x_j x_k \equiv z_{ijk}\), whereby constraints are imposed to ensure equivalence between the higher-order terms and auxiliary representations. Detailed information about reducing higher-order terms can be found in \cite{Kaseb2024power}.

\subsection{Solve QUBO using quantum/digital annealers}\label{sec:qubosolving}
\noindent The resulting QUBO problem can then be iteratively solved on quantum/digital annealers, where each iteration minimizes the objective function to find the optimal bitstring that best satisfies the original model. If the value of \cref{eq:qubo} drops below a certain user-defined threshold \( \epsilon \), the so-computed \( \hat{x} \) is accepted as the solution value. Otherwise, the base value \( x^0 \) is updated according to \eqref{eq:update}, and the minimization problem \eqref{eq:qubo} is solved again with the updated QUBO matrix $Q$ until convergence is reached.

\subsection{\( \Delta x \) Update}\label{sec:deltaupdate}
\noindent In each iteration, two sets of binary variables, \( \mathbf{x}^{(\text{it}-1)} \) and \( \mathbf{x}^{(\text{it}-2)} \), are stored, representing the bitstrings from the previous and second-to-last iterations, respectively. These bitstrings guide an adaptive update of \( \Delta x \). For instance, if \( x_{0}^{(\text{it}-2)} = 1 \) and \( x_{1}^{(\text{it}-2)} = 0 \), followed by \( x_{0}^{(\text{it}-1)} = 1 \) and \( x_{1}^{(\text{it}-1)} = 0 \), and then a transition to \( x_{0}^{(\text{it})} = 0 \) and \( x_{1}^{(\text{it})} = 1 \) in the current iteration, this indicates that \( \hat{x} \) increased over the last two iterations before reversing direction. Such a pattern suggests instability, where reducing \( \Delta x \) can help mitigate oscillations.

\section{Application to Linear Energy System Problems: 2D Steady Conductive Heat Transfer}
\label{sec:HT}
\noindent Determining the temperature distribution in a solid domain is a fundamental problem in heat transfer analysis. For steady-state conductive heat transfer, the governing equation follows from Fourier's law together with conservation of energy. For a two-dimensional domain, the heat conduction equation is:
\begin{equation}
\nabla \cdot (k \nabla T) = 0 ,
\end{equation}
where \(T\) is the temperature field and \(k\) is the thermal conductivity of the material. Assuming homogeneous material properties with constant thermal conductivity \(k=1\) and no internal heat sources or sinks, the equation simplifies to Laplace's equation:
\begin{equation}
\frac{\partial^2 T}{\partial x^2} + \frac{\partial^2 T}{\partial y^2} = 0.
\end{equation}

To numerically approximate the temperature distribution, the plate is discretized into an \(K \times K\) grid, resulting in \(K^2\) interior nodes with unknown temperatures. Using a finite difference approximation of the Laplacian operator, the temperature at each interior node \((m,n)\) satisfies:
\begin{equation}
T_{m,n} = \frac{1}{4}\left(T_{m+1,n} + T_{m-1,n} + T_{m,n+1} + T_{m,n-1}\right),
\end{equation}
which states that, at steady state, the temperature at each node equals the average of its four neighbors. Note that applying boundary conditions modifies the equations for nodes adjacent to the domain boundaries. Collecting the equations for all nodes yields a linear system of the form:
\begin{equation}
\mathbf{A}\mathbf{T} = \mathbf{b}, \quad \mathbf{T}, \mathbf{b} \in \mathbb{R}^{K^2}, \ \mathbf{A} \in \mathbb{R}^{{K^2} \times {K^2}}.
\end{equation}
where \(\mathbf{T} = \begin{bmatrix} T_1 & T_2 & \dots & T_{K^2} \end{bmatrix}^T\) is the vector of unknown nodal temperatures, \(\mathbf{A}\) is the coefficient matrix determined by the finite difference discretization, and \(\mathbf{b} = \begin{bmatrix} b_1 & b_2 & \dots & b_{K^2} \end{bmatrix}^T\) is the contributions from the boundary conditions.

The problem is to identify the complete \(\mathbf{T}\) given \(\mathbf{A}\) and \(\mathbf{b}\). Reformulating as an optimization problem, we seek to minimize the mismatch:
\begin{equation}
\mathbf{A}\mathbf{T} - \mathbf{b}=0.
\end{equation}

The continuous unknown temperatures are discretized as:
\begin{equation}
\hat{T}_{j} = T_{j}^{0} + {\Delta T}_{j} \left( x_{j,0}^T - x_{j,1}^T \right) \quad j \in \{0,1,\dots,K^2\}.
\end{equation}

Squaring the residual function ensures non-negative penalization, leading to the problem Hamiltonian \(H(x)\) containing \(K^2\) linear equations. The target residual is zero. Therefore, all equations should ideally vanish independently. The squared residual function is given by:
\begin{equation}
\min_{x \in \{0,1\}^{2K^2}} \sum_{i=1}^{K^2} \left( \sum_{j=1}^{K^2} A_{i,j} \hat{T}_{j} - b_i\right)^2.
\end{equation}

\begin{figure}[h]
    \centering
\includegraphics[width=3.7in]{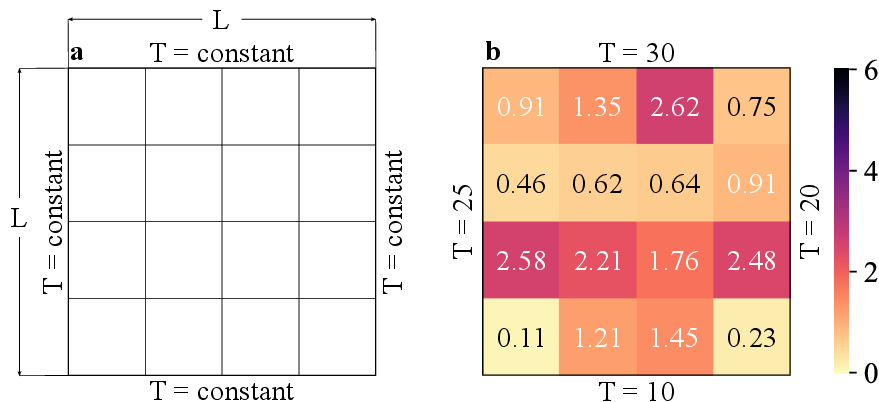}
    \caption{(a) A 2D plate of length $L$ discretized into a $4 \times 4$ grid. (b) Relative error (in percentage) of the 2D steady-state heat conduction solution compared to a classical linear solver. Constant temperature boundary conditions are applied at all four edges. For example, $0.91\%$ indicates that the absolute difference between the estimated and reference values, normalized by the reference value and multiplied by $100$, equals $0.91\%$.}
    \label{fig:ht}
\end{figure}

Solving the resulting linear system yields the equilibrium temperature distribution across the plate, as shown in \cref{fig:ht}a. We solve (11) for a 2D plate of length \(L\) discretized using a \(4 \times 4\) grid. For this case, the matrix \(\mathbf{A}\) has the sparse structure:
\begin{equation}
A =
\begin{bmatrix}
-4 & 1 & 0 & 0 & 1 & 0 & \dots & 0 \\
1 & -4 & 1 & 0 & 0 & 1 & \dots & 0 \\
0 & 1 & -4 & 1 & 0 & 0 & \dots & 0 \\
0 & 0 & 1 & -4 & 0 & 0 & \dots & 0 \\
1 & 0 & 0 & 0 & -4 & 1 & \dots & 0 \\
0 & 1 & 0 & 0 & 1 & -4 & \dots & 0 \\
\vdots & \vdots & \vdots & \vdots & \vdots & \vdots & \ddots & 1 \\
0 & 0 & 0 & 0 & 0 & 0 & 1 & -4
\end{bmatrix}.
\end{equation}

The experiments are conducted using quantum annealing (QA). The estimated temperature vector \(\mathbf{T}\) is compared with reference values obtained using a direct dense linear solver based on LU decomposition with partial pivoting, implemented through NumPy’s \texttt{linalg.solve()} routine, which internally calls the LAPACK \texttt{dgesv} algorithm. The corresponding deviations are shown in \cref{fig:ht}b. The results indicate that all estimated temperatures exhibit a maximum relative error of less than \(2.7\%\).

\section{Application to Linear Power System Problems: Power System Parameter Identification}
\label{sec:LinProblem}
\noindent Power system parameter identification is a critical task in state estimation and system monitoring, enabling the determination of unknown network parameters using available bus voltage and current measurements. This problem is fundamentally governed by Ohm's law, which forms a linear system of equations with complex variables. For an \(N\)-bus power system, the relationship between injected currents \(\mathbf{I}\) and bus voltages \(\mathbf{V}\) is expressed using the system admittance matrix \(\mathbf{Y}\) as:
\begin{equation} \label{eq:ohm}
    \mathbf{I} = \mathbf{Y}\mathbf{V}, \quad \mathbf{I}, \mathbf{V} \in \mathbb{C}^N, \ \mathbf{Y} \in \mathbb{C}^{N \times N}.
\end{equation}

Given measurements of \(\mathbf{V}\) and \(\mathbf{I}\), the unknown elements of \(\mathbf{Y}\) can be estimated by solving this linear system, typically using least squares or other optimization techniques.

The problem is to identify the complete \(\mathbf{Y}\) using bus voltage and current measurements. Reformulating this as an optimization problem, we seek to minimize the mismatch:
\begin{equation} \label{eq:y-matrix}
    \mathbf{I} - \mathbf{Y}\mathbf{V} = 0, 
\end{equation}
where \(\mathbf{Y}=\mathbf{G}+j\mathbf{B}\) and the unknown variables are the conductance \(\boldsymbol{G}\) and susceptance \(\boldsymbol{B}\) of the power system. Expanding \cref{eq:y-matrix} in rectangular form yields:
\begin{equation} \label{eq:y-matrix-expanded}
    \left(\text{Re}(\mathbf{I}) + j \text{Im}(\mathbf{I})\right) - \left(\boldsymbol{G} + j\boldsymbol{B}\right) \left(\boldsymbol{\mu} + j\boldsymbol{\omega}\right) = 0, 
\end{equation} 
where \(\boldsymbol{\mu}\) and \(\boldsymbol{\omega}\) are the real and imaginary components of \(\mathbf{V}\), respectively. 

The continuous unknowns are discretized as:
\begin{subequations} \label{eq:gb-increment}
    \small\begin{align}
         \hat{G}_{ik} & = G_{ik}^0 + {\Delta G}_{ik} \left( x_{ik,0}^G - x_{ik,1}^G\right), \quad i,k \in \{0,1,\dots,N\}, \label{eq:g-increment}\\
        \hat{B}_{ik} & = B_{ik}^0 + {\Delta B}_{ik} \left( x_{ik,0}^B - x_{ik,1}^B\right), \quad i,k \in \{0,1,\dots,N\}. \label{eq:b-increment}
    \end{align}
\end{subequations}

Squaring the residual function ensures non-negative penalization, leading to the problem Hamiltonian \(H(\mathbf{x})\), including real and imaginary components. The target residual is zero, i.e., \(0 + j0\). Hence, both components should independently vanish. The squared residual function is:
\begin{subequations} \label{eq:leq-hamiltonian}
    \begin{align} 
        \min_{x \in \{0,1\}^{4N}} \sum_{k=1}^{N}\left(\text{Re}(I_k)-\sum_{i}^{N}\mu_i \hat{G}_{ki}+\sum_{i}^{N}\omega_i\hat{B}_{ki}\right)^2 + \sum_{k=1}^{N} \left(\text{Im}(I_k) - \sum_{i}^{N} \omega_i \hat{G}_{ki} - \sum_{i}^{N} \mu_i \hat{B}_{ki} \right)^2.
    \end{align}
\end{subequations}

\begin{figure}[h]
\centering
\includegraphics[width=2in]{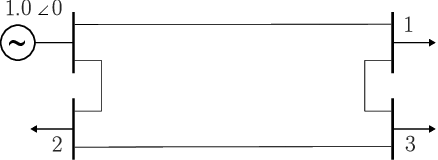}
\caption{Schematic representation of a 4-bus test system, including a slack bus and three load buses (buses 1-3).}
\label{fig:4-bus}
\end{figure}

\begin{figure}[h]
    \centering
\includegraphics[width=3.5in]{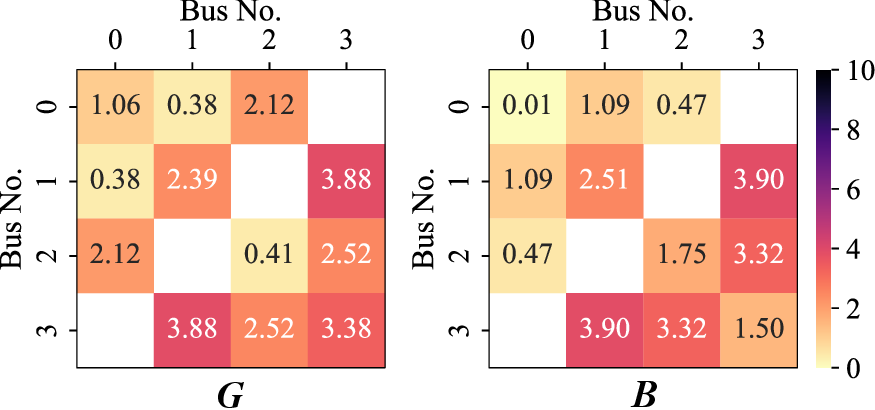}
    \caption{\(\boldsymbol{G}\) and \(\boldsymbol{B}\) relative error (in percentage) representation comparing with reference values from the Power System Test Cases library for the 4-bus test system. For instance, $G_{0,0}=1.06\%$ means that the absolute difference between the estimated and reference values, normalized by the reference value and multiplied by $100$, is $1.06\%$.}
    \label{fig:y}
\end{figure}

We solve \cref{eq:leq-hamiltonian} for a standard 4-bus test system~\cite{Grainger1994}, consisting of one slack bus and three load buses, as shown in \cref{fig:4-bus}. The goal is to determine line properties \(\mathbf{Y}=\mathbf{G}+j\mathbf{B}\) given voltages and current measurements for this test system. The experiments are conducted using QA. The estimated \(\boldsymbol{G}\) and \(\boldsymbol{B}\) matrices are compared against the reference values available in the Power System Test Cases library for the standard 4-bus test system. The corresponding deviations are shown in \cref{fig:y}. The results indicate that all estimated parameters have a maximum relative error of less than $3.9\%$.


\section{Application to Nonlinear Power System Problems: Power Flow Analysis}
\label{sec:NonLinProblem}
\noindent PF analysis is a fundamental task in power systems, aiming to determine complex bus voltages that satisfy the power balance equations at each bus in an \(N\)-bus system:
\begin{subequations} \label{eq:pq-mismatch} 
    \small \begin{align} 
        \mathbf{P} - \mathbf{P}^G + \mathbf{P}^D  = 0, \quad \mathbf{P}, \mathbf{P}^G, \mathbf{P}^D \in \mathbb{C}^N, \label{eq:p-balance}\\
        \mathbf{Q}- \mathbf{Q}^G + \mathbf{Q}^D  = 0, \quad \mathbf{Q}, \mathbf{Q}^G, \mathbf{Q}^D \in \mathbb{C}^N, \label{eq:q-balance}
    \end{align} \normalsize
\end{subequations}
where \(\mathbf{P}\) and \(\mathbf{Q}\) are the net active and reactive power injections, respectively, \(\mathbf{P}^G\) and \(\mathbf{Q}^G\) are generated active and reactive power, respectively, and \(\mathbf{P}^D\) and \(\mathbf{Q}^D\) are consumed active and reactive power, respectively. The formulation in \cref{eq:pq-mismatch} is classically solved using iterative numerical methods, such as the NR solver.

The net power injections in rectangular coordinates form a nonlinear system of equations with complex variables and can be expressed at each bus \(i \in \{0,1,\dots,N\}\) as:
\begin{subequations} \label{eq:pq-sum_expanded}
    \small \begin{align}
        P_k & = \sum_{k=1}^{N} \hat{\mu}_i G_{ik} \hat{\mu}_k + \hat{\omega}_i G_{ik} \hat{\omega}_k + \hat{\omega}_i B_{ik} \hat{\mu}_k - \hat{\mu}_i B_{ik} \hat{\omega}_k , \label{eq:p-sum_expanded}\\ 
        Q_k & = \sum_{k=1}^{N} \hat{\omega}_i G_{ik} \hat{\mu}_k - \hat{\mu}_i G_{ik} \hat{\omega}_k - \hat{\mu}_i B_{ik} \hat{\mu}_k - \hat{\omega}_i B_{ik} \hat{\omega}_k , \label{eq:q-sum_expanded}
    \end{align}
\end{subequations}
where \(\hat{\mu}_i\) and \(\hat{\omega}_i\) are discretized unknown variables:
\begin{subequations} \label{eq:muomega-increment}
\small \begin{align}
    \hat{\mu}_i & = \mu_i^0 + \Delta \mu_i \left( x_{i,0}^\mu - x_{i,1}^\mu \right), \quad i \in \{0,1,\dots,N\} , \label{eq:mu-increment}\\
    \hat{\omega}_i & = \omega_i^0 + \Delta \omega_i \left( x_{i,0}^\omega - x_{i,1}^\omega \right), \quad i \in \{0,1,\dots,N\}. \label{eq:omega-increment}
\end{align}
\end{subequations}

The objective function for PF analysis, \(F^2(\hat{\mathbf{\mu}}, \hat{\mathbf{\omega}})\), is derived by squaring the mismatch equations in \cref{eq:pq-mismatch}, which includes quadratic terms, and hence, \(F^2(\cdot)\) contains linear, quadratic, cubic, and fourth-order terms:
\begin{equation} \label{eq:pf-hamiltonian}
    \min_{x \in \{0,1\}^{4N}} \sum_{i=1}^{N} \left(P_i - P_i^G + P_i^D\right)^2 + \left(Q_i - Q_i^G + Q_i^D\right)^2.
\end{equation}

\begin{figure}[h]
    \centering
\includegraphics[width=2.7in]{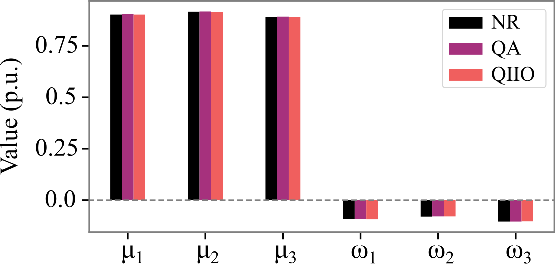}
    \caption{Performance comparison of QA and QIIO with NR in approximating the real ($\mu_i$) and imaginary ($\omega_i$) components of bus voltages at the load buses (buses 1-3) in the 4-bus test system.}
    \label{fig:pf}
\end{figure}

We reduce higher-order terms to obtain a QUBO representation and then solve \cref{eq:pf-hamiltonian} for the 4-bus test system. The experiments are conducted using QIIO and QA. The results are compared with those obtained from the NR solver in Pandapower. \cref{fig:pf} presents the computed bus complex voltage ($\mu_i + j\omega_i$) obtained from the NR, QA, and QIIO solvers. The slack bus ($i=0$), with known $\mu_0=1$ and $\omega_0=0$, is omitted from \cref{fig:pf}. The results show strong agreement across all solvers, with the maximum relative error remaining below $0.3\%$ for both QIIO and QA. Note that QA converged in 98 iterations (average QPU time 0.015 seconds per iteration), while QIIO required only 24 iterations (0.06 seconds per iteration) to reach the same tolerance of $\epsilon = 10^{-4}$, both excluding communication overhead.

\section{Discussion and Conclusion}
\noindent We propose a novel combinatorial optimization framework that reformulates continuous energy and power system problems into a format executable on emerging Ising machines. The framework is demonstrated on both linear problems (2D steady conductive heat transfer and power system parameter identification) and a nonlinear problem (PF analysis). The maximum relative error compared to the reference values for 2D steady conductive heat transfer is 2.7\%, for power system parameter identification is 3.9\%, and for PF analysis is 0.3\%. These results suggest that the proposed framework is applicable to a broad range of problems that can be formulated as root-finding tasks.

The proposed framework has the potential to address unresolved challenges in energy and power systems for several reasons. It accommodates both normal and complex variables. It also enables the dynamic reconfiguration of known and unknown variables. For example, while this work estimates the admittance matrix in \( \mathbf{I} = \mathbf{Y}\mathbf{V} \), the framework remains applicable when solving for voltages \(\mathbf{V}\) or currents \(\mathbf{I}\). The framework can also be extended to incorporate inequality constraints by introducing slack variables, enabling applications, such as OPF~\cite{Kaseb2025enhanced}. Furthermore, the framework allows a partitioned formulation, in which only a subset of the system of equations is included. As demonstrated in~\cite{Kaseb2024power}, this approach retains accuracy while improving computational efficiency. Our previous study on PF analysis showed that the proposed approach can handle large-scale, ill-conditioned cases that conventional numerical methods struggle with~\cite{Kaseb2025enhanced}. As quantum hardware continues to evolve toward commercial viability, the framework also offers the potential for faster convergence and real-time optimization~\cite{Kaseb2025QHIL}. While the current implementation focuses on Ising machines, the developed problem Hamiltonian can also be solved using Quantum Approximate Optimization Algorithm (QAOA) on gate-based quantum hardware, as demonstrated in~\cite{Kaseb2025qaoa}. However, the problem sizes currently realizable with QAOA are significantly smaller than those achievable with quantum and digital annealers.

Several aspects require further investigation in future work to improve efficiency. The discretization scheme employed is straightforward yet effective and can be further refined using preferred number series, such as the 1-2-5 series, to enable finer updates per iteration and reduce the number of steps required for convergence. That said, the framework relies on discretizing continuous variables, which directly affects solution accuracy. A systematic analysis of how sensitive the results are to discretization choices and how these choices influence stability should be conducted in future work. The step size \( \Delta x \) is also adaptively updated based on performance in previous iterations. Future improvements could incorporate reinforcement learning approaches that enable an agent to learn optimal step-size updates. Furthermore, the validation and extension of the proposed framework to more complex power system problems will be the subject of future research.

\section*{Acknowledgment}
\noindent This work is part of the DATALESs project, with project no. 482.20.602, jointly financed by the Netherlands Organization for Scientific Research (NWO) and the National Natural Science Foundation of China (NSFC). This work used the Dutch national e-infrastructure, with support from the SURF Cooperative under grant no. EINF-6569. The authors acknowledge the J\"ulich Supercomputing Centre for providing computing time on the D-Wave Advantage\texttrademark\, System JUPSI through the Jülich UNified Infrastructure for Quantum computing (JUNIQ). The research received support from the Center of Excellence RAISE, which is funded by the European Union's Horizon 2020-Research and Innovation Framework Programme (H2020-INFRAEDI-2019-1) under grant agreement no. 951733. The authors would like to further thank Fujitsu Technology Solutions for providing access to the QIIO software and to Matthieu Parizy for his support.

\printcredits

\bibliographystyle{unsrt}
\bibliography{cas-refs}

\end{document}